\documentstyle[aps,prd,psfig,eqsecnum,twocolumn]{revtex}

\begin{document}
\title{New Constraints from High Redshift Supernovae and Lensing Statistics
upon Scalar Field Cosmologies}
\author{Ioav Waga$^{1}$ and Joshua A. Frieman$^{2,3}$}
\address{$^{1}$Universidade Federal do Rio de Janeiro, 
Instituto de F\'\i sica\\
Rio de Janeiro, RJ, 21945-970, Brazil\\
$^{2}$NASA/Fermilab Astrophysics Center, \\
Fermi National Accelerator Laboratory\\
PO Box 500, Batavia IL 60510, USA \\
$^{3}$Department of Astronomy and Astrophysics\\
University of Chicago, Chicago, IL 60637 USA \\ 
}
\date{\today }
\maketitle

\begin{abstract}
We explore the implications of gravitationally lensed QSOs
and high-redshift SNe Ia observations for
spatially flat cosmological models in which a 
classically evolving scalar field currently dominates the energy density of the
Universe. We consider two representative scalar field potentials that give rise
to effective decaying $\Lambda $ (``quintessence'') models:
pseudo-Nambu-Goldstone bosons ($V(\phi )=M^4\left( 1+\cos \left( \phi /f\right)
\right) $) and an inverse power-law potential ($V(\phi )=M^{4+\alpha }\phi
^{-\alpha }$). We show that a large region of parameter space 
is consistent with current data if $\Omega_{m0} > 0.15$. On the other hand, 
a higher lower bound for the matter density parameter suggested by large-scale
galaxy flows, $\Omega_{m0} > 0.3$, considerably reduces the allowed 
parameter space, forcing the scalar field behavior to approach that of a 
cosmological constant. 
\end{abstract}

\section{Introduction}

Recent observations of type Ia supernovae (SNe Ia) at high redshift 
suggest that the
expansion of the Universe is accelerating \cite{rie98,per99}:
these calibrated `standard' candles appear fainter than would be 
expected if the expansion were slowing due to gravity.
While concerns about systematic errors (such as   
possible evolution of the source population and grey dust) remain, 
the current evidence indicates that the high-redshift 
supernovae appear fainter because, at fixed redshift, they are 
at larger distances. According to the Friedmann equation, 
${{\ddot{a}}/a} =-(4\pi G/3)(\rho + 3p)$, accelerated expansion 
requires a dominant component with either negative energy density, which is 
physically inadmissible, or effective negative
pressure. Dark energy, dynamical-$\Lambda $ (dynamical vacuum energy), or
quintessence are different names that have been used to denote  
 this component. A cosmological
constant, with $p_\Lambda = - \rho_\Lambda$, is the simplest possibility.

Recent studies incorporating new CMB data \cite{sdlk99,boom99} confirm previous
analyses suggesting a large value for the total density parameter, $\Omega
_{total}>0.4$, and favor a nearly flat Universe ($\Omega_{total} = 1$). 
A different set of
observations \cite{omega} now unambiguously point to low values 
for the matter density parameter, $\Omega
_{m0}=0.3\pm 0.1$. In 
combination, these two results provide independent evidence for the conventional 
interpretation of the SNe Ia results and 
strongly support a spatially flat cosmology with $\Omega _{m0}\sim 0.3$ and a
dark energy component with $\Omega _{X}\sim 0.7$. These models 
are also theoretically appealing since a dark energy component that is 
homogeneous on small scales (20--30 $h^{-1\text{ }}$ Mpc) reconciles 
the spatial flatness predicted by 
inflation with the sub-critical value of $\Omega _{m0}$ \cite{pee84}.

The cosmological constant has been introduced several times in modern cosmology to
reconcile theory with observations \cite{revlam} and subsequently 
discarded when improved data or interpretation showed it was not needed.
However, it may be 
that the ``genie'' will now remain forever out of the bottle \cite
{zeldov}. Although current cosmological observations favor a
cosmological constant, there is as yet no explanation why its value is 50 to
120 orders of magnitude below the naive estimates of quantum field theory. One of
the original motivations for introducing the idea of a dynamical $\Lambda$-term
was to alleviate this problem. There are also observational motivations for considering dynamical-$\Lambda$ as opposed to constant-$\Lambda$ models.
For instance, the COBE-normalized amplitude of
the mass power spectrum is in general lower in a dynamical-$\Lambda$ model than
in a constant-$\Lambda $ one, in accordance with observations
\cite{cdf97}. Further, since 
distances are smaller (for fixed $z$ and $\Omega_{m0}$), constraints from the 
statistics of lensed QSOs 
are weaker in dynamical-$\Lambda$ models\cite{rat92,wm99,fhsw95}.

\section{Scalar Field Models}

A number of models with a dynamical $\Lambda $ have been discussed in the
literature \cite{rat88,fhsw95,cal98,dyn,xfluid,scalar}. We report here new
constraints from gravitational
lensing statistics and high-z SNe Ia on two representative scalar field
potentials that give rise to effective decaying $\Lambda $ models: 
pseudo-Nambu-Goldstone bosons (PNGB), with potential  
of the form 
$V(\phi )=M^4\left( 1+\cos \left( \phi /f\right) \right) $, and
inverse power-law models, $V(\phi )=M^{4+\alpha }\phi ^{-\alpha }$. 
These two models are chosen to be representative of the range of dynamical 
behavior of scalar field `quintessence' models. In the PNGB model, 
the scalar field at early times is frozen and therefore acts as a 
cosmological constant; at late times, the field becomes dynamical, 
eventually oscillating about the potential minimum, and the 
large-scale equation of state approaches that of non-relativistic matter ($p=0$). 
The power-law 
model, on the other hand, exhibits ``tracker'' solutions \cite{rat88,track}: 
at high redshift, the scalar 
field equation of state is close to that of non-relativistic matter, and 
at late times it approaches that of the cosmological constant.

Let us consider first the motivation for the PNGB model. All ``quintessence'' 
models involve a scalar field with ultra-low effective mass. 
In quantum field
theory, such ultra-low-mass scalars are not {\it generically} natural:
radiative corrections generate large mass renormalizations at each order of
perturbation theory. To incorporate ultra-light scalars into particle
physics, their small masses should be at least `technically' natural, that
is, protected by symmetries, such that when the small masses are set to
zero, they cannot be generated in any order of perturbation theory, owing to
the restrictive symmetry.  Pseudo-Nambu-Goldstone bosons
(PNGBs) are the simplest way to have naturally ultra--low mass, spin--$0$
particles. These 
models are characterized by two mass scales, a spontaneous
symmetry breaking scale $f$ (at which the effective Lagrangian still retains
the symmetry) and an explicit breaking scale $M$ (at which the effective
Lagrangian contains the explicit symmetry breaking term). 
In order to act approximately like a
cosmological constant at recent epochs with $\Omega _{\phi }\sim 1$, the
potential energy density should be of order the critical density, $M^{4}\sim
3H_{0}^{2}m_{Pl}^{2}/8\pi $, or $M\simeq 3\times 10^{-3}h^{1/2}$ eV. As usual
we set $V=0$ at the minimum of the potential by the assumption that the fundamental
vacuum energy of the Universe is zero -- for reasons not yet understood.
Further, since observations indicate an accelerated expansion, at present
the field kinetic energy must be relatively small compared to its
potential energy. This implies that the motion of the field is still
(nearly) overdamped, that is, $\sqrt{|V^{\prime \prime }(\phi _{0})|}
\lesssim 3H_{0}=5\times 10^{-33}h$ eV, i.e., that the PNGB is ultra-light. 
The two conditions above imply 
that $f\sim m_{Pl}\simeq 10^{19}$ GeV. 
Note that $M\sim 10^{-3}$ eV is close to the neutrino mass scale for
the MSW solution to the solar neutrino problem, and $f\sim m_{Pl}\simeq
10^{19}$ GeV, the Planck scale. Since these scales have a plausible origin
in particle physics models, we may have an explanation for the `coincidence'
that the vacuum energy is dynamically important at the present epoch \cite
{fhsw95,fhw,fukyan}. Moreover, the small mass $m_\phi \sim M^2/f$ is technically natural.

Next consider the inverse power-law model: this potential gives rise to 
attractor (tracking) solutions. If $\rho_\phi$ and $\rho_B$ denote 
the mean scalar and dominant background (radiation or matter) densities, then 
if $\rho _{\phi }\ll \rho _{B}$, the
following `tracker' relationship is satisfied: $\rho _{\phi }^{TR}\sim a^{3(\gamma
_{B}-\gamma _{\phi }^{TR})}\rho _{B}$, where $\gamma _{\phi }^{TR}=\gamma
_{B}\;\alpha /(2+\alpha )<\gamma _{B}$ \cite{rat88,track}. Here, 
$a(t)$ is the cosmic scale factor, and $\gamma _{B} = (p_B+\rho_B)/\rho_B$ 
denotes the adiabatic index of the
background ($\gamma _{B}=4/3$ during the radiation-dominated era and $
\gamma _{B}=1$ during the matter-dominated epoch (MDE)). If the scalar field is in 
the tracker solution, 
its energy density decreases more slowly than the background energy
density, and the field eventually begins to dominate the dynamics of the
expansion. If the field is on track during the MDE, its effective adiabatic
index is less than unity---its effective pressure $p_\phi =
(\dot{\phi}^2/2)-V(\phi)$ is 
negative. This condition by itself does not guarantee accelerated expansion: 
the field must have sufficiently negative pressure and a sufficiently 
large energy density such that the total effective adiabatic index (of the 
field plus the matter) is less than 2/3. Moreoever, 
for inverse power-law potentials, at late times $\Omega _{\phi }\rightarrow 1$, such
that when the growing $\Omega _{\phi }$ starts to become appreciable, 
$\gamma _{\phi }$ deviates from the above tracking value, decreasing toward
the value $\gamma _{\phi }\rightarrow 0$. Thus, even if $\alpha >4$, such
that initially $\gamma_{\phi} = 
\gamma _{\phi }^{TR}>2/3$ in the MDE, when the field begins to dominate the 
energy density  
and $\gamma _{\phi }$ decreases, the Universe will enter a
phase of accelerated expansion. If $\Omega _{m0}$ and $\alpha$ are sufficiently 
small, this will happen before the present time. For inverse power-law
potentials, the two
conditions $\Omega _{\phi 0}\sim 1$ and the preponderance of the field
potential energy over its kinetic energy (the condition for negative pressure)  
imply $M\sim 10^{\frac{27 \alpha-12}{\alpha + 4}}$ eV and $
\phi _{0}\sim m_{Pl}$. Since $\phi _{0}\sim m_{Pl}$, quantum gravitational 
corrections to the potential may be important and could invalidate this picture 
\cite{carrol98}.

In the very early Universe, in order to successfully achieve tracking, the
scalar 
field energy density must be smaller than the radiation energy
density. If, in addition, $\rho_\phi$ is smaller than the initial value of the tracking
energy density, the field will remain frozen until they have comparable
magnitude; at that point, 
the field starts to follow the tracking solution. On the other hand, if $\rho_\phi$
is larger than the initial value of the tracking energy density, the field will
enter a phase of kinetic energy domination ($\gamma _{\phi }\sim 2$); this 
causes $\rho _{\phi }$ to decrease rapidly ($\rho _{\phi }\propto a^{-6}$), overshooting the
tracker solution \cite{track}. Subsequently, as in the case above, the field 
is frozen and later begins to follow the tracking solution when its 
energy density becomes comparable to 
the tracking energy density. In either case, 
there is always a phase before tracking during which the field is frozen. 
Consequently, an important variable 
is the value of the field energy density when it freezes.
For instance, is it smaller or larger than $\rho _{eq}$, the mean energy density
at the epoch of radiation-matter equality? Did the field have time to
completely achieve tracking or not? In fact, the exact constraints imposed by
cosmological tests on the parameter space of this model depend upon this condition.

In a previous study \cite{fw98}, we numerically evolved the scalar field 
equations of motion forward from the epoch of matter-radiation equality, assuming 
the field is initially frozen, ${\dot \phi}(t_{eq}) = 0$. 
In this case, depending on the values of $\alpha $ and $\Omega
_{m0}$, it may happen that the field does not have 
time to reach the tracking solution before the present. 
In general, if $\Omega _{m0}$ is large, we observe that at the present $\gamma
_{\phi }$ is still growing away from its initial value $\gamma _{\phi }=0$. On 
the other hand, if $\Omega _{m0}$ is sufficiently low, $
\gamma _{\phi }$ will reach a maximum value (not necessarily the tracking
value) at some point 
in the past and at the present time will be decreasing to the 
value $\gamma _{\phi }\rightarrow 0$. 
Here we follow a different approach. In our numerical computation we now start
the evolution of the scalar field during the radiation dominated epoch and
assume that it is on track early in the evolution of the Universe.\footnote{
In fact this is true only if $\alpha$ is not close to zero. The case $
\alpha=0$ is equivalent to a cosmological constant, and the field remains
frozen always.} When $\rho _{\phi }$ becomes non-negligible compared to the 
matter density, 
$\gamma _{\phi} $ starts to decrease toward zero.
Recently, constraints from high-z SNe Ia on power-law potentials with the 
field rolling with this set of initial conditions were obtained by Podariu and
Ratra\cite{pod99}. We complement their analysis by including the lensing
constraints as well. In the next section we show using the scalar field
equations  
that present data prefer low values of $\alpha $. We also update and expand the 
observational constraints on the PNGB models \cite{fw98}.

\section{Observational Constraints}

In the following we briefly outline our main assumptions for lensing and
supernovae analysis. Our approach for lensing statistics
is based on Refs: \cite{koch93,koch96} and is described in more detail in \cite{wm99}. To
perform the statistical analysis we consider data from the HST Snapshot survey
(498 highly luminous quasars (HLQ)), the Crampton survey (43 HLQ), the Yee
survey (37 HLQ), the ESO/Liege survey (61 HLQ), the HST GO observations (17
HLQ), the CFA survey (102 HLQ) , and the NOT survey (104 HLQ) \cite{mao}. We
consider a total of 862 ($z>1$) highly luminous optical quasars plus 5
lenses. The lens galaxies are modeled as singular isothermal spheres (SIS), and
we consider lensing only by early-type galaxies, since they are expected to 
dominate the lens population. We assume a conserved comoving
number density of lenses, $n=n_{0}(1+z)^{3}$, and a  Schechter form for the early
type galaxy population, $n_{0}=\int_{0}^{\infty }n_{*}\left( \frac{L}{L^{*}}
\right) ^{\alpha }\exp \left( -\frac{L}{L^{*}}\right) \frac{dL}{L^{*}}$ ,
with $n_{*}=0.61\times 10^{-2} h^{3}\mbox{Mpc}^{-3}$ and $\alpha =-1.0$ \cite
{lov94}. We assume that the luminosity satisfies the Faber-Jackson relation 
\cite{fab76}, ${L}/{L^{*}}=({\sigma _{||}}/{\sigma _{||}^{*}})^{\gamma }$,
with $\gamma =4$. Since the lensing optical 
depth depends upon the fourth power of the velocity dispersion of an $L^*$
galaxy, a correct estimate 
of this quantity is crucial for strong lensing calculations. The image
angular separation is also very sensitive to $\sigma _{||}^{*}$:  
larger velocities give rise to larger image separations. In our likelihood analysis we
take into account the observed image separation of the lensed quasars and 
adopt the value $\sigma _{||}^{*}=225$ km/s, which gives 
the best fit to the observed image separations \cite{koch96}. 

\begin{figure} 
\hspace*{0.in}
\psfig{file=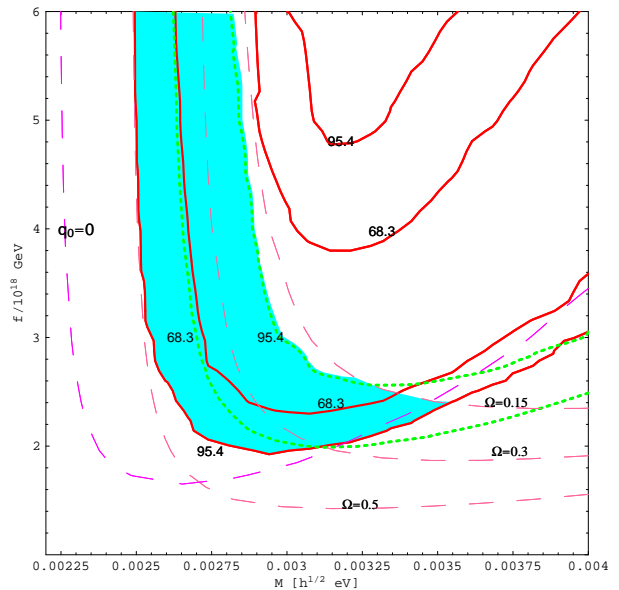,height=8 cm,width= 8 cm} 
\vspace*{0.4in}
\caption{Contours  of constant likelihood ($95.4\%$ and $68.3\%$) arising from
lensing statistics (the region above and to the right of the short dashed
curves is excluded) and type Ia supernovae (solid curves) 
are shown for the PNGB model. Also shown are contours of constant $\Omega_{m0}$
and the limit for present acceleration, $q_0=0$. The shaded region shows the parameter space
allowed at $95\%$ C.L. by the lensing, SNe, and cluster observations.} 
\end{figure} 

For SIS, the total lensing 
optical depth can be expressed analytically, $\tau
(z_{S})=\frac{F}{30}(d_{A}(0,z_{S})(1+z_{S}))^{3}(cH_{0}^{-1})^{-3}$, where 
$z_{S}$ is the source redshift, $d_{A}(0,z_{S})$ is its angular diameter
distance, and $F=16\pi ^{3}n(cH_{0}^{-1})^{3}(\sigma _{||}^{*}/c)^{4}\Gamma
(1+\alpha +4/\gamma )\simeq 0.026$ measures the effectiveness of the lens in
producing multiple images \cite{tog}. 
We correct the optical depth for the effects of magnification bias and 
include the selection 
function due to finite angular resolution and dynamic range \cite{koch93,koch96,wm99}. 
We assume a mean optical extinction of $\Delta m$ =$0.5$ mag, 
as suggested by Falco {\it et al.} \cite{fal98}: this makes the lensing 
statistics for optically selected quasars consistent with the results for radio
sources, for which there is no extinction. When applied to spatially flat cosmological 
constant models, our approach yields the upper bounds $\Omega _{\Lambda
}\lesssim 0.76$ (at $2\sigma$) and $\Omega _{\Lambda
}\lesssim 0.61$ (at $1\sigma$), with a best-fit value of 
$\Omega_{\Lambda}\simeq 0.39$. Recent 
statistical analyses using both HLQ and radio sources slightly 
tighten these constraints on a cosmological constant 
\cite{fal98}.
A combined (optical+radio) lensing analysis
for dynamical-$\Lambda$ models is still in progress; qualitatively, 
we expect this to tighten the lensing constraints below by approximately 
$1\sigma$. 

For the SNe Ia analysis \cite{fw98}, we consider the latest published data from  the High-z Supernovae Search
Team \cite{rie98}\cite{obs}. We use the 27 low-redshift and 10 high-redshift SNe Ia (including SN97ck) reported
in Riess {\it et al.} \cite{rie98} and consider data with the MLCS \cite
{rie96,rie98} method applied to the supernovae light curves. Following a
procedure similar to that described in Riess {\it et al.}\cite{rie98}, we
determine the cosmological parameters through a $\chi ^{2}$ minimization, 
neglecting the unphysical region $\Omega _{m0}<0$.

\begin{figure} 
\hspace*{0.in}
\psfig{file=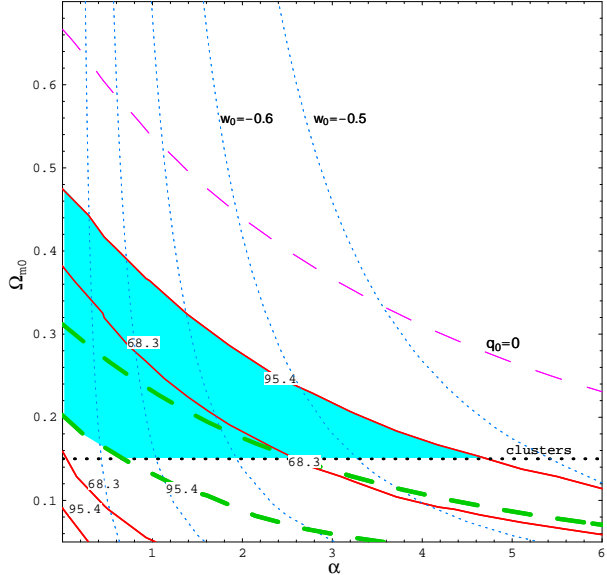,height= 8 cm,width= 8 cm} 
\vspace*{0.4in}
\caption{Contours of constant likelihood ($95.4\%$ and $68.3\%$) arising from
lensing statistics (the region below the thick dashed curves is excluded) and 
type Ia supernovae (solid curves) are shown for the inverse power-law model. 
Also shown is the lower bound $\Omega_{m0}=0.15$ from clusters and curves of 
constant present equation of state $w_0 = p_{\phi 0}/\rho_{\phi 0}$. The shaded region shows the parameter space
allowed at $95\%$ C.L. by the lensing, SNe, and cluster observations.} 
\end{figure}

In Fig. 1 we show the $95.4\%$ and $68.3\%$ C. L. limits from lensing (short
dashed contours) and the SNe Ia data (solid curves) on the parameters $f$ and $M$ of the
PNGB potential. As in \cite{fw98}, these limits apply to models with the
initial condition $\frac{4\sqrt{\pi }\phi (t_{i})}{m_{Pl}}=1.5$ and $\frac{
d\phi }{dt}(t_{i})=0$, with $t_{i}=10^{-5}t_{0}$ ; for other choices, the
bounding contours would shift by small amounts in the $f-M$ plane. We also
plot some contours of constant $\Omega _{m0}$ (dashed) and the curve 
$q_{0}=0$ (long dashed contour)
as a function of the parameters $f$ and $M$. The allowed region (shown by the shaded
area in Fig. 1)
is limited by the lensing and SNe Ia $95.4\%$ C. L. contours and also by 
the constraint $\Omega _{m0}>0.15$, which we interpret as $2\sigma$ lower bound 
from observations of galaxy clusters.
The data clearly favors accelerated expansion (the region above the $q_{0}=0$
curve) but curiously 
there is a small region in the parameter space, close to the
point where the $\Omega_{m0}=0.15$ and the Sne Ia $2\sigma$ curves cross,
where the Universe is {\it not} in accelerated expansion by the present
time. This small area disappears if we adopt the tighter constraint 
$\Omega _{m0}>0.3$. We note that the bulk of the $2\sigma$-allowed parameter space, 
where the lensing and SNe contours are nearly vertical, corresponds to the 
scalar field being nearly frozen, i.e., in this region the model is 
degenerate with a cosmological constant.

In Fig. 2  we show the $95.4\%$ and $68.3\%$ C. L. limits from lensing 
(thick dashed contours) and the SNe Ia data (solid curves) 
on the parameters $\alpha $ and $\Omega_{m0}$ of the inverse power-law
potential. The horizontal dotted line shows a lower bound on the matter 
density inferred from the dynamics of galaxy clusters,  $\Omega _{m0}=0.15$.
We also show contours of the present 
equation of state $w_{0}=\gamma_0 -1$ (thin dotted curves) and the 
curve $q_{0}=0$ (long dashed curve). At $95.4\%$ confidence, the SNe Ia and 
$\Omega_{m0}$ constraints require 
$\alpha <5$ and $w _0 < -0.5$; the latter bound agrees roughly with the constraint 
obtained by assuming a time-independent equation of state 
\cite{wm99}, an approximation sometimes used 
for the inverse power-law model. 
We also observe that the lensing constraints on the model parameters are weak, 
constraining only low values of $\Omega_{m0}$ and $\alpha$. We remark, however, that 
they are consistent with the SNe Ia constraints. We can tighten the constraints
on the 
equation of state if we consider a higher value for the $\Omega _{m0}$ lower
bound. 
For instance, if we adopt $\Omega _{m0}>0.3$, as suggested in \cite{zeh99}, 
we obtain $w_0 < -0.67$ and $\alpha < 1.8$. In both models, a larger lower
bound on $\Omega_{m0}$ pushes the scalar field behavior toward that of the cosmological constant
($w=-1$). 

\section{Conclusion}

A consensus is beginning to emerge that we live in a nearly flat,
low-matter-density Universe with 
$\Omega_{m0} \sim 0.3$ and a dark energy, negative-pressure component with 
$\Omega_X \sim 0.7$.
The nature of this dark energy component is still not well understood; 
further developments will require deeper understanding of fundamental physics
as well as improved observational tests to measure the equation of state at 
recent epochs,  
$w(t)$, and determine if it is distinguishable from that of the cosmological constant \cite{ch99}. 
Classical scalar field models provide a simple dynamical framework for posing 
these questions.
In this paper we analyzed two representative scalar field models, 
the PNGB and power-law potentials, which span the range of expected dynamical 
behavior. 
The inverse power-law model displays tracking solutions \cite{track} which  
allow the scalar field to start from a wide set of initial conditions. We
showed that 
current data favors a small value of the parameter, $\alpha < 5$. This may 
be a problem for these models: in Refs:\cite{track} it was shown that, 
starting from the equipartition condition after inflation, it is necessary to
have 
$\alpha>5$ for the field to begin tracking before matter-radiation
equality. Since the observational 
constraints indicate that tracking could only be achieved 
(if at all) at more recent times, it is not clear what theoretical advantage,
in terms of alleviating the `cosmic coincidence' problem,  
is gained by the tracking solution.
Although well motivated from the particle physics viewpoint, the PNGB model
is strongly constrained by the SNe Ia and lensing data. 
Finally, as noted above, these two models predict radically different 
futures for the Universe. In the inverse power law model, the expansion  
will continue accelerating and approach de Sitter space. In the PNGB model, 
the present epoch of acceleration may be brief, followed by a return to 
what is effectively matter-dominated evolution.

\acknowledgments

We would like to deeply thank Luca Amendola, Robert Caldwell, Cindy Ng, Franco 
Ochionero, Silviu Podariu and Bharat Ratra for several useful discussions that 
helped us to improve this work. This work was supported by the Brazilian 
agencies CNPq and FAPERJ and by the DOE and NASA Grant NAG5-7092 at Fermilab.


\begin{references}
\bibitem{rie98} A. G. Riess {\it et al.},Astron. J. {\bf 116 }, 1009  (1998); P. M. Garnavich {\it et al.}, Astrophys. J. {\bf 509}, 74 (1998).
\bibitem{per99}S. Perlmutter {\it et al.}, Astrophys. J. {\bf 517}, 565, (1999).
\bibitem{sdlk99}S. Dodelson and L. Knox, astro-ph/9909454.
\bibitem{boom99}A. Melchiorri {\it et al.}, astro-ph/9911445.
\bibitem{omega} N. A. Bahcall, J. P. Ostriker, S. Perlmutter and P. J. Steinhardt Science, {\bf 284}, 1481 (1999); M. S. Turner, astro-ph/9901109. 
\bibitem{pee84}P.J.E. Peebles, Astrophys. J. {\bf 284}, 439 (1984); M. S. Turner, G.Steigman, L. M. Krauss, Phys. Rev. Lett. {\bf 52}, 2090 (1984).
\bibitem{rat92}B. Ratra and A. Quillen, Mon. Not. R. Astron. Soc., {\bf 259}, 738, (1992); L. F. Bloomfield Torres and I. Waga, Mon. Not. R. Astron. Soc. {\bf 279}, 712 (1996); A. R. Cooray, Astron. Astrophys. {\bf 342}, 353 (1999).
\bibitem{wm99} I. Waga and A. P. M. R. Miceli, Phys. Rev. D {\bf 59}, 103507, (1999).
\bibitem{zeldov} Y. B. Zeldovich, Sov. Phys. Uspekhi {\bf 11 }, 381 (1968).
\bibitem{revlam}S. Weinberg, Rev. Mod. Phys. {\bf 61}, 1 (1989); S. M. Carroll, W. H. Press and E. L. Turner, Annu. Rev. Astron. Astrophys.,{\bf 30}, 499 (1992); J. Frieman, in ``Third
Paris Cosmology Colloquium'', eds. H. J. de Vega \& N. Sanchez (World
Scientific, 1995); V. Sahni and A. Starobinsky, astro-ph/9904398.
\bibitem{fhw} J. A. Frieman, C. T. Hill, and R. Watkins, Phys. Rev. D {\bf 46},
1226 (1992).
\bibitem{fhsw95}J. A. Frieman, C. T. Hill, A. Stebbins and I. Waga, Phys. Rev. Lett. {\bf 75}, 2077 (1995).
\bibitem{fukyan}M. Fukugita and T. Yanagida, preprint YITP/K-1098 (1995).
\bibitem{cdf97}K. Coble, S. Dodelson, and J. A. Frieman, Phys. Rev. D {\bf 55}, 1851 (1997).
\bibitem{fw98}J. A. Frieman and I. Waga, Phys. Rev. D {\bf 57}, 4642 (1998).
\bibitem{cal98}R. R. Caldwell, R. Dave, and P.J. Steinhardt, Phys. Rev. Lett. {\bf 80}, 1582 (1998).
\bibitem{rat88}B. Ratra and P. J. E. Peebles, Phys. Rev. D {\bf 37}, 3407 (1988); P. J. E. Peebles and B. Ratra, Astrophys. J. {\bf 325}, L17 (1988).
\bibitem{dyn}M. Ozer and M. O. Taha, Nucl. Phys. {\bf B287}, 776 (1987); K. Freese {\it et al.}, Nucl. Phys. {\bf B287}, 797 (1987); M. Reuter and C. Wetterich, Phys. Lett. {\bf B188}, 38 (1987); W. Chen and Y. S. Wu, Phys. Rev. D {\bf 41}, 695 (1990); J. C. Carvalho, J. A. S. Lima and I. Waga, Phys. Rev. D {\bf 46}, 2404, (1992); I. Waga, Astrophys. J. {\bf 414}, 436 (1993); V. Silveira and I. Waga, Phys. Rev. D {\bf 50}, 4890 (1994); V. Silveira and I. Waga, Phys. Rev. D {\bf 56}, 4625 (1997); J. M. Overduin and F. I. Cooperstock, Phys. Rev. D {\bf 58}, 043506, (1998).
\bibitem{xfluid}A. Vilenkin, Phys. Rev. Lett., {\bf 53}, 1016 (1984); J. N. Fry, Phys. Lett. {\bf B158}, 211 (1985); 
H. A. Feldman and A. E. Evrard, Int. J. Mod. Phys. {\bf D2}, 113 (1993); 
J. Stelmach and M. P. Dabrowski, Nucl. Phys.  {\bf B406}, 471 (1993); H. Martel, Astrophys. J. {\bf 445}, 537 (1995); L. M. A. Bittencourt, P. Laguna, R. A. Matzner, hep-ph/9612350; M. Kamionkowski and N. Toumbas, Phys. Rev. Lett., {\bf 77}, 587 (1997); D. N. Spergel and U. L. Pen, Astrophys. J. {\bf 491}, L67 (1997); T. Chiba, N. Sugiyama, and T. Nakamura, Mon. Not. R. Astron. Soc. {\bf 289}, 5 (1997); M. S. Turner and M. White, Phys. Rev. D {\bf 56}, R4439 (1997); M. White, Astrophys. J. {\bf 506}, 485 (1998); W. Hu, Astrophys. J. {\bf 506}, 495 (1998); S. Perlmutter, M.S. Turner and M. White, Phys. Rev. Lett. {\bf 83}, 670 (1999); M. Bucher and D. Spergel, Phys. Rev. D {\bf 60}, 043505, (1999); L. Wang, R. R. Caldwell, J. P. Ostriker and P. J. Steinhardt, astro-ph/9901388. 
\bibitem{scalar}C. Wetterich, Nuclear Physics {\bf B302}, 668 (1988); V. Sahni, H. A. Feldman and A. Stebbins, Astrophys. J. {\bf 385}, 1 (1992); P. T. P. Viana and A. R. Liddle, Phys. Rev. D {\bf 57}, 674 (1998); P. Ferreira and M. Joyce, Phys. Rev. Lett., {\bf 79}, 4740 (1997); Phys. Rev. D {\bf  58}, 023503 (1998); A. Liddle and R. Scherrer, Phys. Rev D {\bf  59}, 023509 (1999); J. Uzan, Phys. Rev D {\bf 59}, 123510 (1999); P. Bin\'etruy, Phys. Rev D {\bf 60}, 063502 (1999); A. Masiero, M. Pietroni and F. Rosati, hep-ph/9905346; L. Amendola, astro-ph/9906073 ; astro-ph/9908023; A. Albrecht and C. Skordis, astro-ph/9908085;  V. Sahni and L. Wang , astro-ph/9910097;
\bibitem{track}I. Zlatev, L. Wang and P. J. Steinhardt, Phys. Rev. Lett. {\bf 82}, 896 (1999); P. J. Steinhardt, L. Wang and I. Zlatev, Phys. Rev D {\bf 59}, 123504 (1999).
\bibitem{carrol98}S. M. Carroll, Phys. Rev. Lett {\bf 81}, 3067 (1998); C. Kolda and D. H. Lyth, Phys. Lett {\bf B458}, 197 (1999); K. Choi, hep-ph/9912218.
\bibitem{pod99} S. Podariu and B. Ratra, astro-ph/9910527.
\bibitem{mao}D. Maoz {\it et al.}, Astrophys. J. {\bf 409}, 28 (1993); D. Crampton ,R. D. McClure, and J. M. Fletcher, {\it ibid.} {\bf 392}, 23 (1992); H. K. C. Yee , A. V. Filipenko, and D. H. Tang, {\it A. J.} {\bf 105}, 7 (1993); J. Surdej {\it et al., ibid.} {\bf 105}, 2064 (1993); E. E. Falco, in {\it Gravitational Lenses in the Universe}, edited by J. Surdej, D. Fraipont-Caro, E. Gosset, S. Refsdal, and M. Remy (Liege: Univ. Liege), p. 127 (1994); C. S. Kochanek, E. E. Falco, and R. Shild, Astrophys. J. {\bf 452}, 109 (1995); A. O. Jaunsen {\it et al.}, Astron. Astrophys. {\bf 300}, 323 (1995).
\bibitem{tog}E. L. Turner, J. P. Ostriker, J. R. Gott III, Astrophys. J. {\bf 284}, 1, (1984).
\bibitem{fab76} S. M. Faber and R. E. Jackson, Astrophys. J. {\bf 204}, 668 (1976).
\bibitem{sch76}P. Schechter, Astrophys. J. {\bf 203}, 297 (1976).
\bibitem{lov94}J. Loveday, B. A. Peterson, G. Efstathiou and S. J. Maddox, Astrophys. J. {\bf 390}, 338 (1994).
\bibitem{koch93}C. S. Kochanek, Astrophys. J. {\bf 419}, 12 (1993).
\bibitem{koch96}C. S. Kochanek, Astrophys. J. {\bf 466}, 47 (1996).
\bibitem{fal98}E. E. Falco, C. S. Kochanek and J. A. Mu\~noz, Astrophys. J. {\bf 494}, 47 (1998).
\bibitem{obs}The results would not change appreciably if we had considered data from \cite{per99}.
\bibitem{rie96}A. G. Riess, W. H. Press and R. P. Kirchner, Astrophys. J. {\bf 473}, 88 (1996).
\bibitem{zeh99}I. Zehavi and A. Dekel, Nature {\bf 401}, 252 (1999).
\bibitem{ch99} A. R. Cooray and D. Huterer, Astrophys. J. {\bf 513}, L95 (1999); D. Huterer and M. S. Turner, Phys. Rev. D {\bf 60}, 081301 (1999); T. D. Saini, S. Raychaudhury, V. Sahni and A. A. Starobinsky, astro-ph/9910231.
\end{references}
\end{document}